\documentclass[a4paper,fleqn,usenatbib]{mn2e}%[useAMS,usenatbib]{mn2e}

\usepackage{graphicx}
\usepackage{amsmath,amssymb}
\usepackage{tablefootnote}
\makeatletter
\newlength{\abovecaptionskip}%
\makeatother
\usepackage[para,online,flushleft]{threeparttable}
\newcommand{\msun}{\rm{M}_\odot}
\newcommand{\nifs}{^{56}\rm{Ni}}
\newcommand{\snx}{SN\,2010X\,}

\newcommand{\rsun}{\rm{R}_\odot}

\title[Helium giants and RFSNe]%Type Ibc SNe]
{Helium giant stars as progenitors of rapidly fading Type Ibc supernovae}
\author[Kleiser et al.]{Io Kleiser$^{1,2}$\thanks{E-mail:ikleiser@caltech.edu}, Jim Fuller$^{1,2}$, Daniel Kasen$^{3,4,5}$
\\
$^{1}$Department of Astronomy, California Institute of Technology, Pasadena, CA 91125.\\
$^{2}$TAPIR, Mailcode 350-17, California Institute of Technology, Pasadena, CA 91125, USA.\\
$^{3}$Lawrence Berkeley National Laboratory, 1 Cyclotron
  Road, Berkeley, CA 94720.\\
$^{4}$Department of Physics, University of California,
  Berkeley, CA 94720.\\
  $^{5}$Department of Astronomy, University of California,
  Berkeley, CA 94720.\\}

\begin{document}
\maketitle

\begin{abstract}

Type I rapidly fading supernovae (RFSNe) appear to originate from hydrogen-free stars with large radii that produce predominantly shock-cooling light curves, in contrast with more typical $\nifs$-rich SNe Ibc. However, it remains to be determined what types of stars would produce bright shock-cooling light curves without significant contribution from radioactive nickel.
Bare helium stars in the mass range $\sim2-4~\msun$ are known to hydrostatically develop radii as large as 100 $\rsun$ or more due to strong He and C shell burning outside of a core with a sharp density gradient. We produce several such stellar models and demonstrate that, when exploded, these helium giants can naturally produce RFSN light curves.
Since many prototypical SNe Ibc should come from large-radius stars in this mass range as well, we predict that these RFSNe may be distinct from SNe Ibc solely due to the absence of substantial $\nifs$.
\end{abstract}

\begin{keywords}
supernovae: general -- binaries: general -- supernovae: individual: SN 2010X
\end{keywords}

\section{Introduction}

There has been some difficulty in characterizing the stars that give rise to hydrogen-poor rapidly fading supernovae (RFSNe) discovered in recent years. Initially their short light curve rise and fall led to the conclusion that they must be very low-mass, perhaps non-terminal, ejections \citep{kasliwal10}.
However, radiation transport calculations \citep{kleiser14} suggest that some of these objects require relatively large $(\gtrsim 0.3~\msun)$ ejecta masses, implying that $\nifs$ is not the dominant power source, since large ejecta masses with significant nickel content will produce a long-lasting light curve. Observational evidence from more recent RFSNe presented by \citet{drout14} and \citet{shivvers16} also point toward scenarios in which these stars explode inside extended envelopes or winds, suggesting that shock-deposited energy, rather than radioactive nickel, is the primary source of power for the light curve. 

The question of why RFSNe would fail to eject nickel is still unanswered. Perhaps a large CSM-to-ejecta mass ratio could more effectively push the innermost material to fall back onto the remnant through the reverse shock that forms once the ejecta and CSM collide \citep{chevalier89}. Alternatively, as shown previously \citep{macfadyen01}, low explosion energies ($\sim 0.1~\mathrm{B}$) could allow material to fall back, stifling the radioactive material and allowing only shock energy to power the light curve. 
Another possibility is that low shock temperatures may result in very little nickel production in the first place.

There are several possible mechanisms for developing an extended envelope around a hydrogen-free star toward the end of its life. 
A large effective radius (tens to hundreds of $\rsun$) could ensue from dynamical ejection of material in the last few days of the star's life or from heating and expansion of the envelope. One promising avenue for bringing significant mass out to large radii prior to explosion is the mechanism described by \citet{quataert12} in which instabilities in core oxygen burning produce $g$-modes that propagate as $p$-modes through the envelope. Large, thick envelopes could also be the result of common envelope evolution, as discussed by \citet{chevalier12} in the case of SNe IIn. This possibility was invoked speculatively for RFSNe in previous work \citep{kleiser14,kleiser18}.

Here we entertain another possibility, which is more naturally produced in simple stellar evolution calculations. Extended helium red giant stars have been shown to arise from certain binary evolution scenarios and can explode as SNe Ibc
\citep{paczynski71,savonije76,nomoto84,habets86,yoon10, woosley95,dessart18,yoon12,podsiadlowski92,yoon15,eldridge15,yoon17,divine65}. Upsilon sagittari \citep{dudley90,koubsky06} may be an example of such a moderately inflated He star in the midst of case BB mass transfer. If these stars explode, even with a small amount of energy, their light curves could be very bright because of the very extended radius and moderate envelope mass while producing very little $\nifs$. 

These helium stars, typically in the range of 2-4 $\msun$ after stripping, are therefore appealing candidates for RFSNe; they naturally develop very extended radii ($\gtrsim 100~\mathrm{R}_\odot$), and some of them are expected to result in electron-capture SNe (ECSNe) and low-mass iron core-collapse SNe (CCSNe), which should produce very small amounts of $\nifs$ without the need for fallback of material onto the remnant \citep{radice17,muller17,mayle88,sukhbold16}. In this paper, we explore this possibility by running numerical simulations of the evolution of these stars, their explosions, and resulting light curves and spectra.

\section{Methods}

Using MESA version 10000, we model helium stars in the 2-4 $\msun$ mass range using 
a constant mass loss rate of $10^{-3}~\msun/\mathrm{yr}$ after the star has left the Main Sequence and expanded such that its surface temperature has dropped below $\sim 5000~\mathrm{K}$. This threshold is meant to indicate when the star's radius has likely increased enough for Roche lobe overflow. 
Once the H envelope has been removed, the artificial mass loss is shut off. The bare He core is then allowed to evolve until the simulation is stopped. 
We use the default settings for massive stars in MESA, including a ``Dutch" hot wind scheme with scaling factor of 0.8 \citep{glebbeek09}. We use Type 2 opacities and assume solar abundances at the beginning of the simulation. 

We use the final progenitor star model as the input for our 1D hydrodynamics code and run a shock through it after removing the innermost 1.4 $\msun$, assuming this forms the remnant. The explosion energy is chosen by hand and deposited as a thermal bomb by artificially increasing the thermal energy of the innermost few zones. The hydrodynamics code is not coupled to radiation but uses a $\gamma=4/3$ equation of state. 

We feed the output profile into a separate radiation transport code, SEDONA \citep{kasen06}, once the ejecta are roughly free-streaming, as described in \citet{kleiser18}. The implicit assumption is that the ejecta will expand adiabatically and radiation will be trapped until it is homologous. This may not be the case for all objects, e.g. those in which radiation begins escaping before the shock has traversed the entire stellar envelope and interaction is still occurring while the supernova can be observed optically. However, this assumption should be appropriate for many objects, particularly those from intermediate-radius stars. Even in cases where radiation hydrodynamics would be ideal, our results should provide informative rough peak luminosities and decline timescales; the behavior of the rise will not be adequately captured. Therefore, with this simplification, we use SEDONA to calculate time-dependent light curves and spectra for our ejecta profiles beginning about a day after explosion. In some of our light curve calculations, we add $\nifs$ that has essentially a smoothed step-function profile, as described in \citet{kleiser18}.

\section{Results}

We have produced stellar models with varying zero-age Main Sequence mass $M_\mathrm{ZAMS}$ between 12 and 18 $\msun$ such that their bare helium cores lie in the 2-4 $\msun$ range once the hydrogen envelope is removed. For the lower-mass models, the calculation slows dramatically after oxygen core formation due to the overlap of convective regions with thin burning shells. Since the envelope of the star is already quite extended by this time, we stop all models once the radius has settled into a relatively stable state. In models we allowed to run longer, the radius tended to remain constant after this point or increase steadily, but here we show only the evolution up until just after the radius settles following oxygen core formation. The more massive stars are able to evolve further, and we stop them at the point of off-center neon ignition. These stars expand in radius somewhat, although not as much as their lower-mass counterparts.

\begin{figure}
\begin{center}
\includegraphics[width=3.3in]{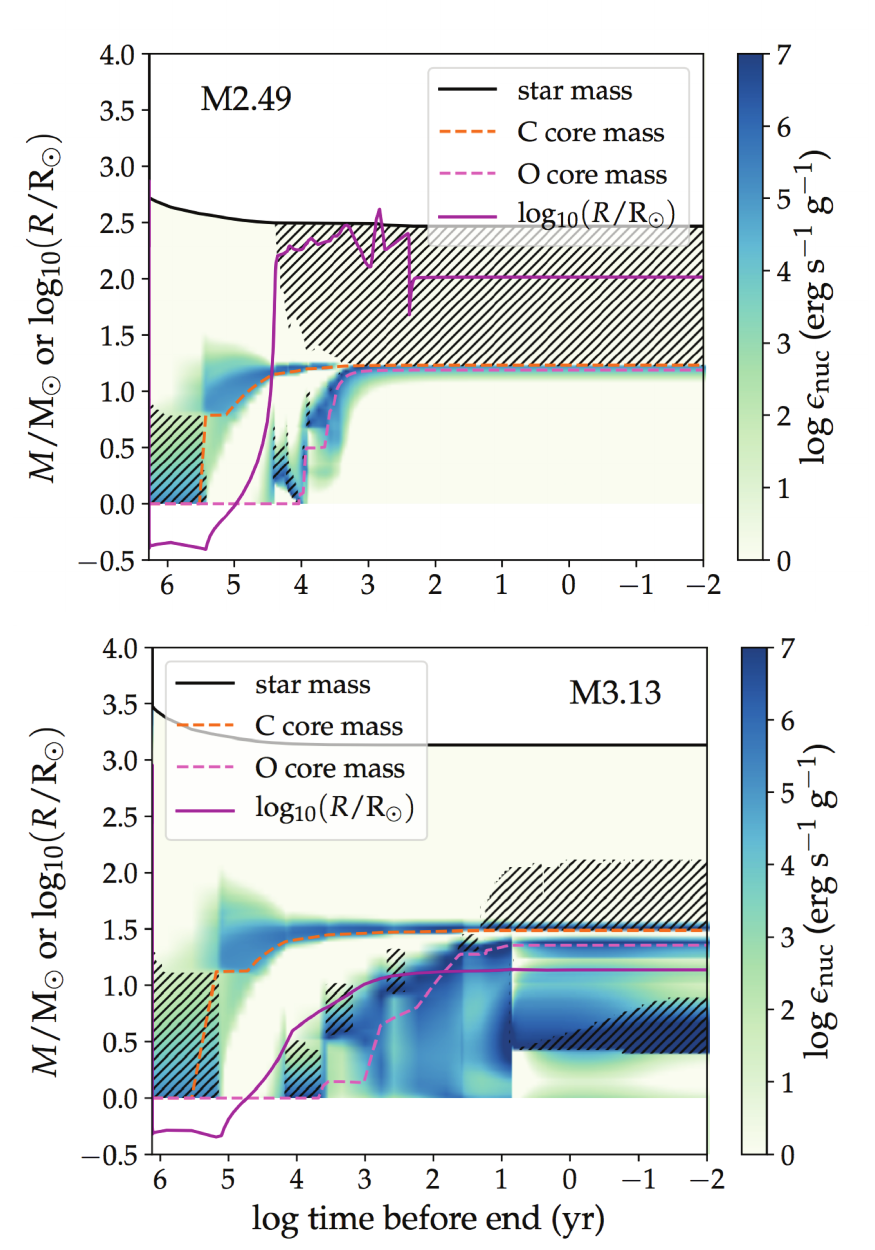}
\end{center}
\caption{Kippenhahn diagrams for a low-mass model ($2.49~\msun$ at the end of artificial mass loss, $M_\mathrm{ZAMS} = 14~\msun$) and high-mass model ($3.13~\msun$, $M_\mathrm{ZAMS} = 16~\msun$). The radius over time is overlain as well, and hatches indicate convective regions. For both stars, the radius expands when the carbon core forms and helium shell burning begins. The radius of the lower-mass star grows dramatically as a convective layer forms at the surface and deepens throughout the envelope, eventually reaching the He and C shell burning regions, which have grown very close to one another. 
This dramatic expansion does not occur for the higher-mass star,
although the evolution and final structure also will depend on the size of the Roche lobe at this point. \label{f:evol_core}}
\end{figure}

We show Kippenhahn diagrams of one low-mass and one high-mass star in Figure \ref{f:evol_core}. Helium shell burning is responsible for the initial expansion of the radius during core carbon burning. As the carbon in the core is exhausted, an oxygen core begins to form and carbon shell burning starts. In the case of a low-mass star, a convective layer develops at the surface and extends inward, which helps inflate the star dramatically. Once the convective envelope penetrates down to the He and C shell burning regions, which are now very narrow and nearly on top of one another, the envelope enters a tumultuous phase and the radius is highly variable before settling into a slower and more steady growth. The differences in behavior between the two types of models is consistent with previous findings \citep[e.g.][]{habets86a,habets86,yoon10}.

The higher-mass stars, by contrast, do not develop a surface convective zone, and the He and C shell burning layers remain separate. Instead, helium shell burning creates a convective shell, but it does not lead to the dramatic expansion seen in the low-mass stars. We can also see that core burning continues in the higher-mass star, whereas the lower-mass star develops a degenerate core. 

\begin{figure}
\begin{center}
\includegraphics[width=3.3in]{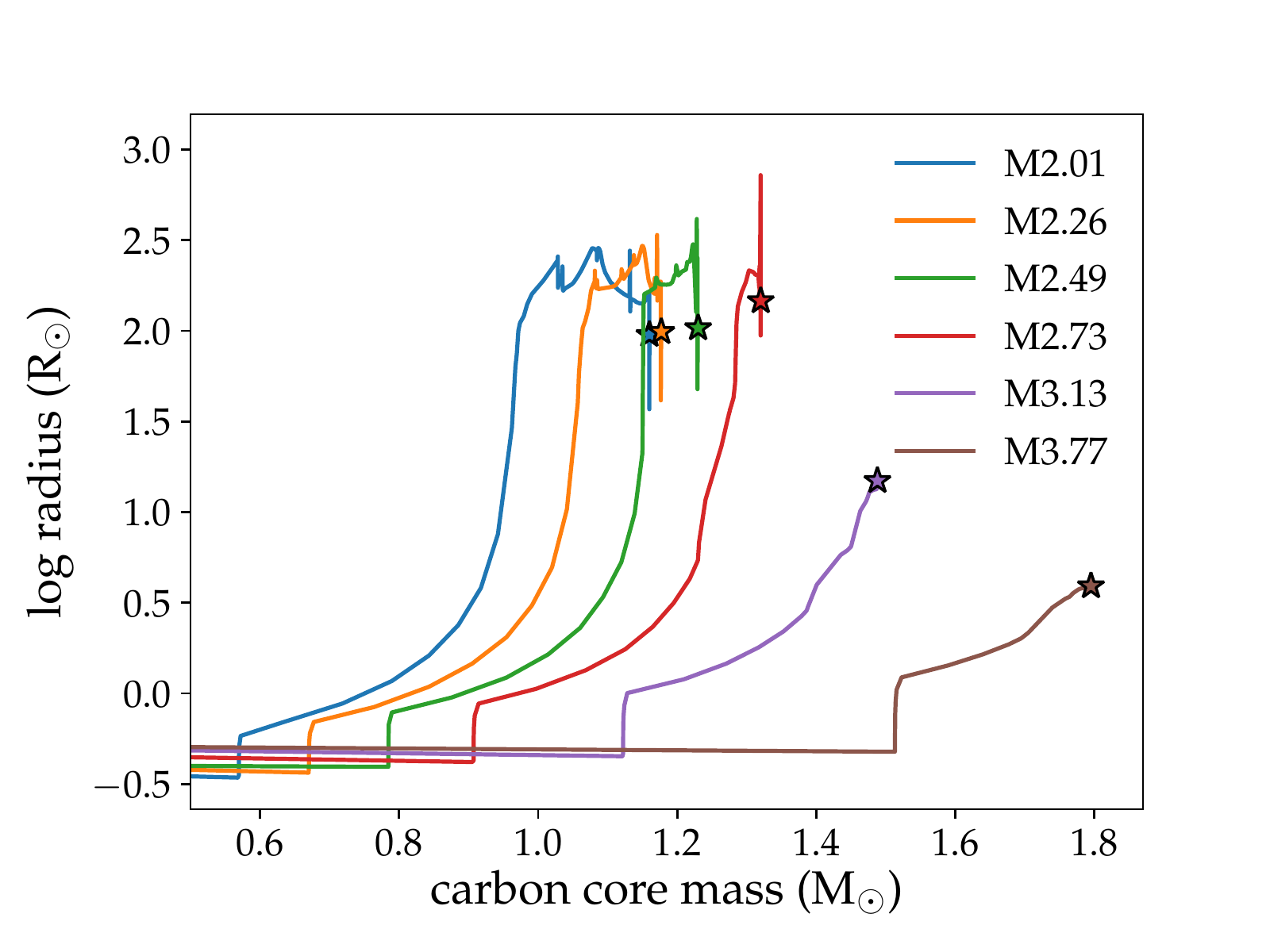}
\end{center}
\caption{Radii as a function of carbon core mass for all stellar models. The radius, which dropped significantly at the onset of mass loss (not shown in this plot),
increases dramatically as shell burning heats the envelope.
 \label{f:radii}}
\end{figure}

Figure \ref{f:radii} shows the photospheric radius as a function of carbon core mass near the end of the star's evolution. 
The lower-mass stars expand dramatically and have some rapid variability before they settle into their final radii. Meanwhile, their higher-mass counterparts expand steadily up to the point of off-center neon ignition, but they only grow to a few solar radii.

\begin{figure}
\begin{center}
\includegraphics[width=3.3in]{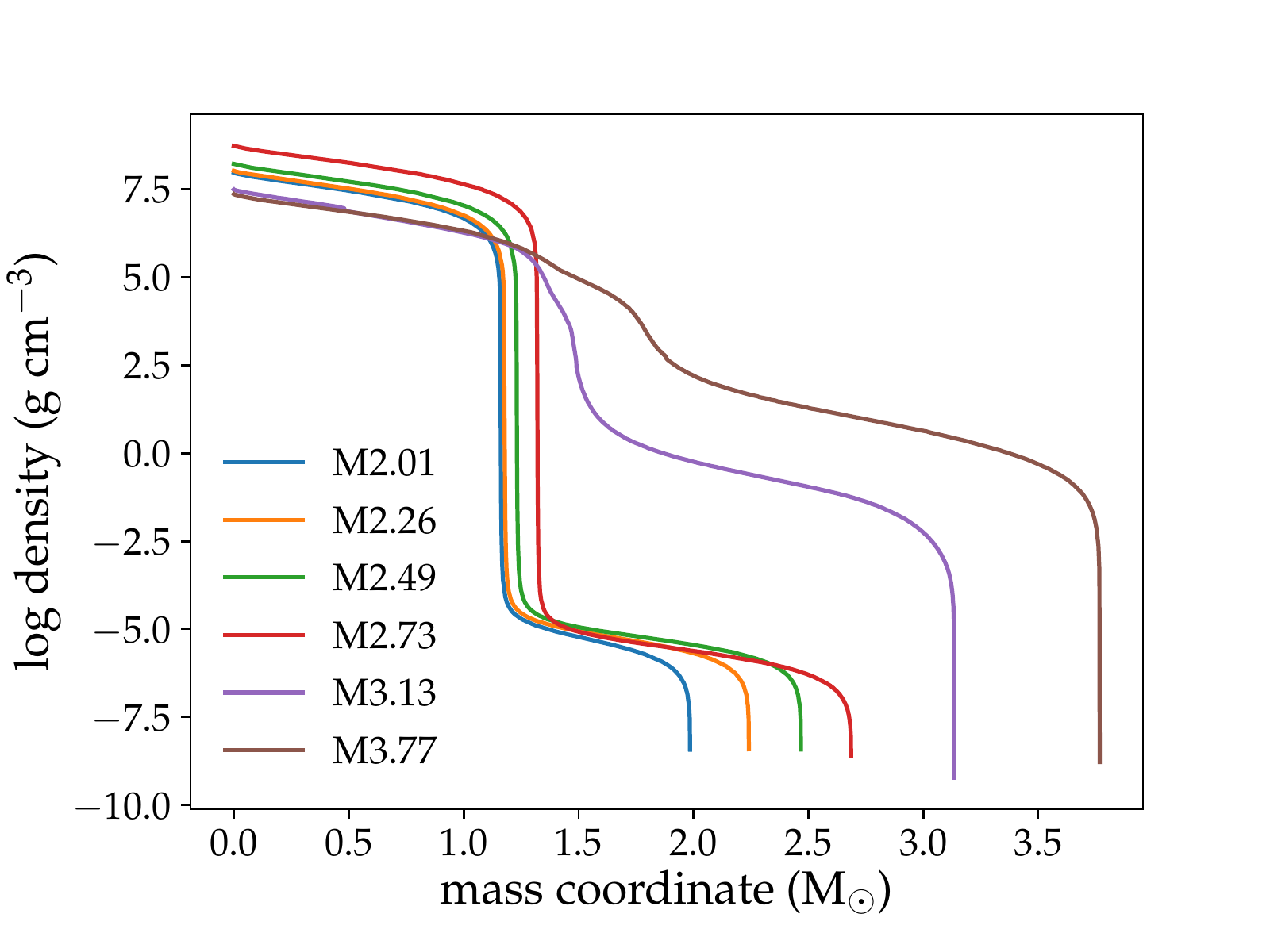}
\end{center}
\caption{Stellar density profiles for final stellar models. The lower-mass stars have steeper density gradients outside their degenerate cores, causing their envelopes to expand to large radii due to helium shell burning. Meanwhile, higher-mass stars have much more even density distributions and much less steep gradients outside the core.
 \label{f:rho_profiles}}
\end{figure}

We show the final stellar density profiles of all models in Figure \ref{f:rho_profiles}. The models with very extended radii show, as discussed in \citet{habets86}, a very steep density gradient outside the core and low-density envelope. Larger-mass models do not feature this density gradient and have a more even distribution of mass.

\begin{figure}
\begin{center}
\includegraphics[width=3.3in]{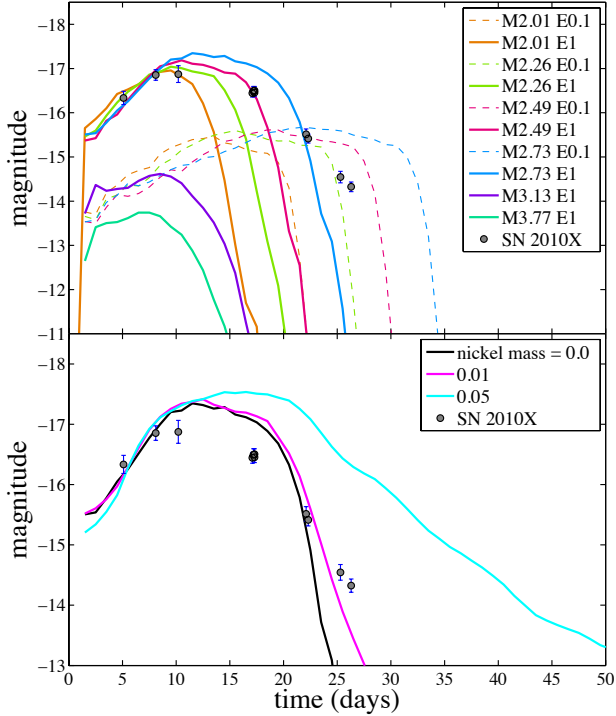}
\end{center}
\caption{Top: Light curves (SDSS $r$-band) for all models. Light curves in the $g$ and $i$ bands track the $r$ band closely in all cases. The peak luminosities and timescales are similar to many known RFSNe, although it is difficult to capture the rapid rise times while maintaining a slow enough decline time, as seen by comparison to \snx. Lower-energy explosions rise more slowly, are slightly more plateau-like, and drop off rapidly. The explosions of low-radius, high-mass helium stars are faint without nickel. Bottom: SDSS $r$-band light curves calculated using the M2.73 E1 explosion model with various amounts of mixed $\nifs$ using the formula in \citet{kleiser18} with the transition from nickel-rich to nickel-poor ejecta spanning $\sim$ 50 out of 200 zones. A small amount of $\nifs$ can produce a tail while most of the peak luminosity still comes from shock cooling.
\label{f:lc_multi}}
\end{figure}

When the stellar models are exploded, shock heating of the envelope converts some of the kinetic energy back into thermal energy. This behavior was explored for toy helium shells added to smaller-radius stars in \citet{kleiser18} but is also well known for giant stars with extended envelopes \citep[see e.g.][]{popov93,woosley95,kasen09}. 
Even for lower-energy ($0.1~\mathrm{B}$) explosions, if the radius is large, then the final thermal energy at $t = 10^5~\mathrm{s}$ can still be significant. 
In Figure \ref{f:lc_multi}, we show light curves from all of our exploded models, some of which are qualitatively similar to RFSNe. 
While rise times are long compared to known objects, peak luminosities and timescales are similar. Low-mass models with large radii produce bright supernovae with relatively short rise times and very rapid decay times. More massive stars with smaller radii produce dim, very short light curves that would be difficult to detect without nickel. The low-mass models we exploded with $0.1~\mathrm{B}$ are moderately bright and more plateau-like than typical RFSNe. \citet{dessart18} also show a shock cooling light curve from a moderately extended helium giant, but it is relatively dim and very short-lived, and they propose this as an early component of SN Ibc light curves rather than as an explanation for RFSN light curves.

Using a similar $\nifs$ abundance profile to the ones explored in \citet{kleiser18}, we add various amounts of mixed nickel to our $15~\msun$ model exploded with 1 B, shown in Figure \ref{f:lc_multi}. If the nickel is more radially mixed (i.e. not centrally concentrated), the peak blends with the shock cooling peak. A radioactive tail is also present, but most of the peak luminosity comes from shock cooling. In a scenario like this, in which an extended helium star explodes with a small amount of highly mixed nickel, it may be difficult to distinguish the light curves from those of a regular SN Ibc that is dominantly nickel-powered. 

\begin{table*}
  \begin{threeparttable}[htp]
    \caption{Stellar and supernova properties.}
    \begin{tabular}{c c c c c c c c c c c c} % centered columns
    \hline\hline %inserts double horizontal lines 
    $M_\mathrm{ZAMS}$ $^\mathrm{a}$ &
$M_\mathrm{ML}$ $^\mathrm{b}$&
$M_\mathrm{final}$ $^\mathrm{c}$&
$M_\mathrm{ccore}$ $^\mathrm{d}$&
$M_\mathrm{ej}$ $^\mathrm{e}$&
$R_\mathrm{ZAMS}$ $^\mathrm{f}$&
$R_\mathrm{ML}$ $^\mathrm{g}$&
$R_\mathrm{final}$ $^\mathrm{h}$&
$E_\mathrm{th}$ $^\mathrm{i}$&
 $m_{r,\mathrm{SN}}$ $^\mathrm{j}$&
 $t_\mathrm{SN}$ $^\mathrm{k}$&
 \\ [0.5ex] % inserts table 
12.0  & 2.18 & 2.01  & 1.08 & 0.61 & 9.59 & 0.352 & 95.2  & 2.54e+49 & -16.96 & 13  \\
13.0  & 2.46 & 2.26  & 1.15 & 0.86 & 10.0 & 0.383 & 98.7  & 3.11e+49 & -17.04 &  16 \\
14.0  & 2.72 & 2.49  & 1.22 & 1.09 & 10.4 & 0.402 & 104  & 3.60e+49 & -17.19 & 18  \\
15.0 & 2.99 & 2.73  & 1.31 & 1.33 & 11.2 & 0.408 & 146  & 5.13e+49 & -17.35 & 21  \\
16.0  & 3.47 & 3.13 & 1.48 & 1.73 & 12.0 & 0.473 & 14.9  & 1.17e+48 & -14.62 & 14  \\
18.0  & 4.24 & 3.77  & 1.79 & 2.37 & 12.2 & 0.548 & 3.91  & 4.11e+47 & -13.74 &  13 \\
\hline %inserts single line 

    \end{tabular}
    \label{t:stars}
    \begin{tablenotes}
    \item [a] Initial or zero-age Main Sequence (ZAMS) mass in $\msun$.
    \item [b] Mass in $\msun$ of bare helium core at end of artificial mass loss.
    \item [c] Mass in $\msun$ of bare helium core at end of calculation.
    \item [d] Carbon core mass in $\msun$ at end of calculation.
    \item [e] Ejecta mass in $\msun$.
    \item [f] Radius (photospheric) on the Main Sequence in $\rsun$.
    \item [g] Radius at end of artificial mass loss in $\rsun$.
    \item [h] Final radius at end of calculation in $\rsun$.
    \item [i] Thermal energy at $10^5$ seconds after the star is exploded with 1 B (erg).
    \item [j] Supernova $r$-band peak absolute magnitude.
    \item [k] Timescale of the supernova (days), from explosion until the r-band luminosity declines by a factor of 2 from peak.
    \end{tablenotes}
  \end{threeparttable}
\end{table*}

\section{Discussion and Conclusions}

We have presented models of bare helium cores from the lower-mass end of massive stars that expand to very large radii toward the end of their lives. This expansion is due to intense shell burning when the core contracts, similar to the mechanism for envelope expansion in hydrogen-rich red giant stars \citep{habets86a,habets86,yoon10}.
The greatest expansion occurs for stars with cores that become very compact with a sharp density gradient above the core during carbon shell burning. The helium stars in these mass ranges would probably end their lives as either iron core-collapse or electron-capture SNe. 

Even without production of radioactive nickel, explosions of stars with such extended radii can produce bright transients, and they qualitatively reproduce the features of some RFSNe discovered in recent years. We have found that lower explosion energies, which may be more relevant for electron-capture SNe, can still yield transients bright enough to detect. He cores from $16~\msun$ stars and above do not develop large radii and would be very difficult to observe without the presence of nickel.

There are several possible explanations for the dearth of nickel available in the ejecta. In previous work \citep{kleiser18}, we considered the fallback of some of the innermost material onto the remnant \citep{macfadyen01,moriya18}. This scenario is unlikely for the helium giants presented here because they have steep density gradients outside their cores and low compactness \citep{oconnor11,sukhbold16}, which more readily allow for neutrino-driven explosions with very little bound material \citep{muller17}. If the star explodes as an electron-capture SN, it is expected to produce very little nickel \citep{nomoto87,miyaji87,mayle88,wanajo09,muller17,poelarends17}. Additionally, core-collapse explosions from iron cores on the lower-mass end should also produce much less nickel than their more massive counterparts \citep{radice17}. \citet{sukhbold16} show iron yields, which can be taken as a proxy for nickel yields, for single stars models; stars with cores comparable to those of helium giants from binaries produce less nickel by a factor of about 10.

An important consideration for the light curves is that the final radii and envelope configurations of these models may evolve beyond what are presented here. If they become even more extended, they could result in even brighter supernovae. Based on preliminary calculations, we speculate that the radii of many of these stars could reach hundreds of $\rsun$. However, the radii may be constrained by companion interaction; if the orbit has not widened enough, the expanding helium star will overflow its Roche lobe and be stripped via case BB mass transfer \citep[see e.g.][]{delgado81,dewi02,tauris17}. Explosions from stars that have undergone such extreme stripping have been explored by \citet{tauris13} and \citet{tauris15}.

We have used a simple mass loss prescription that mimics Case B mass transfer, then allow the star to evolve as though it is a single star. Realistically, a scenario is needed in which the entire hydrogen envelope can be lost to Roche lobe overflow; but at the end of the star's life, the He envelope is allowed to expand without becoming unstable. 
There are several ways to accomplish this. 
One is that the secondary star is more massive than the He star after case B mass transfer. Any subsequent mass loss from the He star due to Roche lobe overflow will cause the orbit to widen, and for some binary configurations the He star may have quite an extended radius at core-collapse.
It will require a more detailed exploration of binary parameters to show what final donor star structures are possible \citep[see e.g.][]{yoon10}. Even if the star overfills the new Roche lobe, the Roche lobe only needs to be $100-200~\rsun$ for the star to produce bright shock-cooling transients. It is also possible that material overflowing the Roche lobe produces a common envelope ejection, which the exploding star might run into as a dynamically ejected shell or wind rather than as an extended hydrostatic envelope.

Alternatively, as described by \citet{dessart18}, the accretor may be originally a lower-mass star that then becomes higher mass once it removes the donor's hydrogen envelope. This star then might evolve faster than its companion and explode as a supernova first, potentially unbinding the binary system and allowing the donor to continue evolving as a single helium star. In this scenario, the helium envelope could then expand unimpeded and possibly reach several hundred $\rsun$. Another way to produce a single helium star would be a common envelope interaction during case B mass transfer, in which a low-mass companion star merges with the core of the donor. The small amount of remaining hydrogen could then be lost via winds during the core helium-burning phase, producing a bare helium core that could expand unimpeded. 

A broader consideration is that many regular SNe Ibc, which are assumed to be mostly radioactively powered, could have a strong shock cooling component \citep{arnett82,bersten14}. As seen in \citet{kleiser18}, adding some nickel to the ejecta can result in double-peaked light curves, but if a small amount is mixed into the ejecta, the peaks may be blended. In fact, the presence of an extended envelope should cause a reverse shock that will cause Rayleigh-Taylor instabilities and mix nickel outward \citep{paxton18}. Therefore the luminosity from any nickel produced in the explosions of these helium red giants would likely blend with the shock cooling component rather than causing a double-peaked light curve. The Rayleigh-Taylor mixing would also change the overall abundance structure, which we do not address here. If blending between a significant shock cooling component and nickel component occurs, then it would be difficult to tease out the contribution of each power source based on the peak luminosity.

\section*{Acknowledgements}

The authors thank Lars Bildsten, Jared Brooks, Sterl Phinney, Evan Kirby, Selma de Mink, Thomas Tauris, and Onno Pols for useful discussion. IK is supported by the DOE NNSA Stockpile Stewardship Graduate Fellowship Program. This research is also funded in part by the Gordon and Betty Moore Foundation through Grant GBMF5076. DK is supported in part by a Department of Energy Office of Nuclear Physics Early Career Award, and by the Director, Office of Energy Research, Office of High Energy and Nuclear Physics, Divisions of Nuclear Physics, of the U.S. Department of Energy under Contract No. DE-AC02-05CH11231.

\bibliographystyle{mn2e} 
\bibliography{paper4}

\end{document}